# Disorder-induced signal filtering with topological metamaterials

*Farzad Zangeneh-Nejad and Romain Fleury*\*

**Disorder, ubiquitously present in realistic structures, is generally thought to disturb the performance of analog wave devices, as it often causes strong multiple scattering effects that largely arrest wave transportation. Contrary to this general view, here, it is shown that, in some wave systems with non-trivial topological character, strong randomness can be highly beneficial, acting as a powerful stimulator to enable desired analog filtering operations. This is achieved in a topological Anderson sonic crystal that, in the regime of dominating randomness, provides a well-defined filtering response characterized by a Lorentzian spectral line-shape. Our theoretical and experimental results, serving as the first realization of topological Anderson insulator phase in acoustics, suggest the striking possibility of achieving specific, non-random analog filtering operations by adding randomness to clean structures.**

The performance of most wave systems, such as lasers, switches, modulators, signal processors etc., is largely impeded by disorder. Even at very low concentration, impurities or geometrical imperfections can cause severe self-interference effects, largely hindering wave propagation and device performance. Recently, inspired by the notion of topological insulators (TIs) in condensed matter systems[1-3], a fascinating solution to mitigate these harmful consequences has been suggested. TIs are insulating phases with non-trivial topological order, supporting edge states that are resilient to certain types and levels of disorder[4-6]. Such an unprecedented property, known as topological protection, promises to alleviate the detrimental effects of disorder on wave propagation, enabling the realization of a large variety of topological analog devices that maintain their original functionality even in the presence of impurities[7-9].

Although TIs have somewhat enhanced the robustness of analog wave systems to disorder, their topological protection is still limited by a phenomenon, known as Anderson localization (AL)[10]. This process, occurring in the regime with dominating randomness, progressively fills the band gap of the TI with disorder-induced localized bulk states, destroying the insulating topological phase and impeding the transportation of the corresponding edge state. Such a behavior seems to be disappointing at first glance, because it implies that even topological wave systems become fragile when the disorder level is high enough to turn the TI into an ordinary insulator. Yet, the mere fact that introducing disorder to a system can induce a topological phase transition is encouraging, because it suggests that the opposite transition might, in principle, be possible. Recently, in a remarkable development[11], it was theoretically demonstrated that some trivial insulators with specific parameters can indeed go through a topological phase transition upon introducing disorder, converting them to TIs with robust conductive states flowing on their boundaries. Soon after, these exotic topological phases, referred to as disorder-induced topological insulators or topological Anderson insulators (TAIs) [12-17], were experimentally observed in different physical platforms[18-20]. Yet, their realization in acoustics has not been reported.

In this work, we provide the first experimental observation of TAI phase in acoustics. We show that the much-sought disorder-induced character of such phases can be leveraged to trigger a well-defined analog filtering operation, namely first-order band-pass filtering[21-25]. Our findings, defying the conventional view that disorder is detrimental to analog wave systems, hold great promises for a large variety of analog devices in which disorder acts as a powerful engine, forcing the system to perform the functionality of interest.

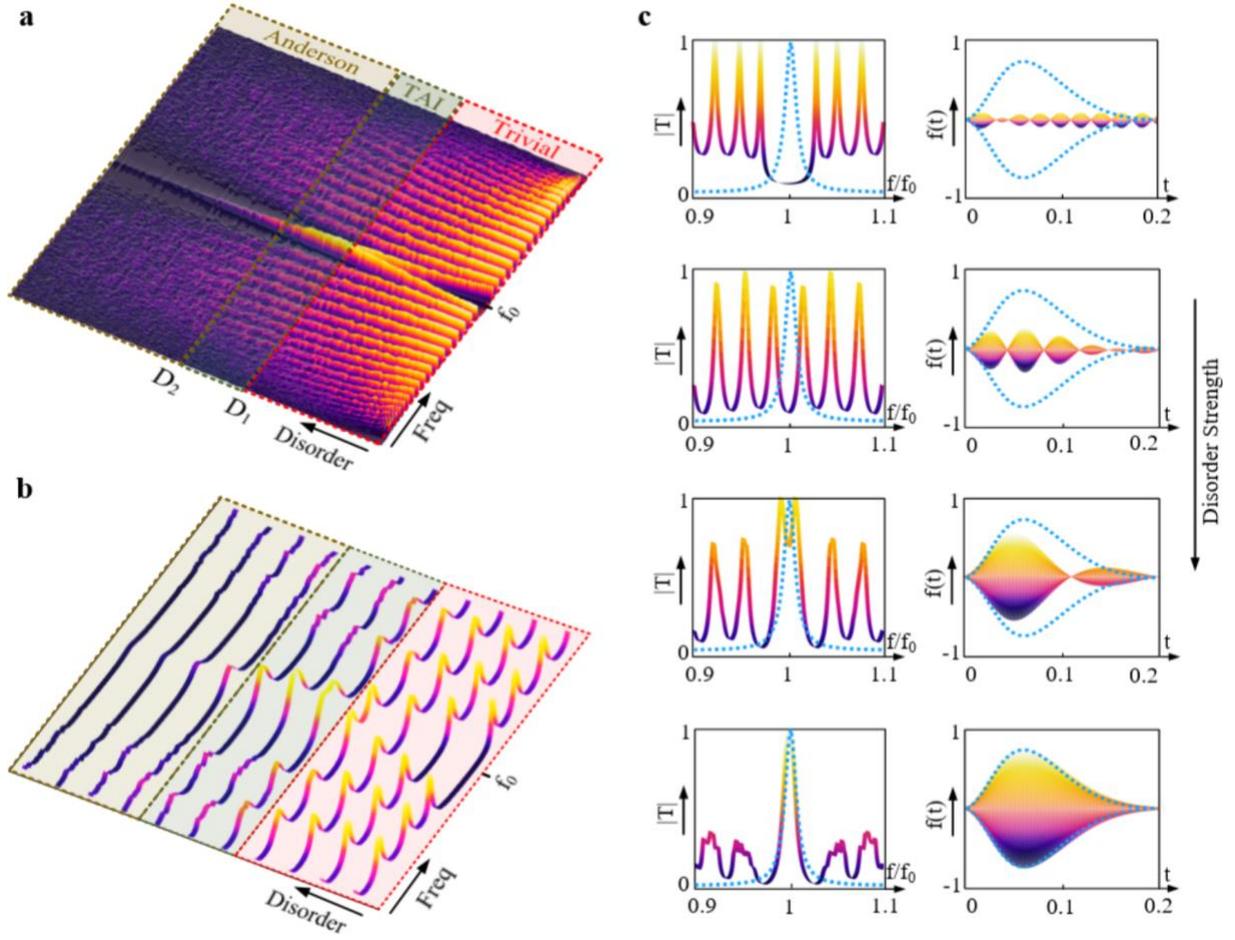

**Figure 1. Disorder-induced analog signal filtering based on topological Anderson insulators**, We consider the disordered version of Su-Schrieffer–Heeger model (SSH) model, described by the tight-binding toy model given in **Equation 1** with the specified parameters, and build a two-port scattering system by coupling a finite crystal piece with 100 unit cells to two external waveguides. **a,** Evolution of the corresponding (averaged) transmission spectrum versus disorder strength. Starting from an ordinary trivial insulator in the clean limit (red region), the system switches into a topological insulator in the regime $D_1 < D_s < D_2$ (TAI regime), characterized by a zero-energy edge state which manifests itself as a resonance peak in the spectrum. For extremely high disorder intensities (yellow region), the transportation is arrested by Anderson localization. **b,** Averaged transmission coefficient of the system for several representative disorder strengths. In the TAI regime (green area), the spectrum exhibits a Lorentzian profile near $f_0$, corresponding to the transfer function of a first-order band-pass filter. **c,** Demonstration of disorder-induced analog filtering. We suppose that the system is excited with a Gaussian-modulated sinusoidal signal and calculate the corresponding transmission coefficient ($T$) and output time signal ($f(t)$), when gradually increasing the disorder strength from zero to the regime of TAI. It is seen that disorder acts like an actuator in our system, applying the desired filtering operation to the input signal.

Let us start with considering the tight-binding toy Hamiltonian of the Su-Schrieffer–Heeger (SSH) chain[1], expressed as

$$H = \sum_n \omega_0 a_n^\dagger a_n + \sum_n K a_{2n-1}^\dagger a_{2n} + \sum_n J a_{2n}^\dagger a_{2n+1} + H.C. \qquad (1)$$

in which $a_n^\dagger$, $a_n$ are creation and annihilation operators for the site $n$, $K$, and $J$ stand, respectively, for the intra-cell and extra-cell coupling coefficients, and $\omega_0 = 1$ is the on-site energy of the atoms. We suppose that the parameters $K$ and $J$ are defined as $K = K_0(1 + 0.5 D_s W)$ and $J = J_0(1 + D_s W)$, in which $K_0 = 0.1, J_0 = 0.09$, $D_s$ is a parameter quantifying the strength of disorder, and $W$ is a site-dependent random number. Since $K > J$ in the clean limit ($D_s \to 0$), the disorder-free system corresponds to a trivial topological phase (BDI class), characterized by a zero winding number[1]. To induce a topological phase transition, we start to increase the intrinsic disorder of the system, now considering the case in which $D_s > 0$. Notice that on average, regardless of $D_s$, the parameter $K$ is always smaller than $J$. Yet, and quite surprisingly, the difference in their standard deviations can create a topological phase transition in a certain range of values for $D_s$ (see Note I Supporting information), leading to an insulator with non-trivial topological index (non-zero winding number). In order to examine such a possibility, we consider a finite two-port scattering system made of 100 unit cells coupled to external waveguides and report in **Figure 1a** the disorder-averaged transmission spectrum versus the parameter $D_s$ (see Note II, Supporting information for the numerical methods). As observed, in the disorder-free case, the spectrum is gapped around $f_0$. This insulating band gap becomes narrower with increasing disorder strength, and is eventually closed at $D_1 = 1.68$. After $D_1$ (and before $D_2 = 3.64$), the band-gap re-opens but, this time, it includes a resonance peak emerging at the center of the gap. This in-gap resonance, corresponding to resonant tunneling through a topological edge state, indicates the non-trivial character of the system under investigation for $D_1 < D_s < D_2$ (TAI regime). If one increases the disorder level further ($D_s > D_2$), the onset of Anderson localization is reached, where all states start to localize in the bulk with decreasing transmission coefficient. In **Figure 1b**, we have plotted the disorder-averaged transmission coefficient for several representative disorder strengths. Notice that, in the TAI regime, the topological mid-gap resonance has a Lorentzian profile, following the general form of $H(\omega) = A/(Bj(\omega - \omega_0) + C)$ in the vicinity of $\omega_0$. Remarkably, $H(\omega)$ corresponds to the transfer function of a first-order band-pass filter. It then follows that our disordered one-dimensional toy model acts as a first-order analog filter, induced by disorder. In particular, while the system under study is performing a definite task, it has a disorder-induced character and an indefinite randomly-drawn geometry, a property that directly stems from the underlying topological Anderson insulator phase.

In order to examine the functionality of the proposed system, we assume that the system is excited with a Gaussian-modulated sinusoidal pulse (see Note III, Supporting information for other forms of excitation signals), and calculate the corresponding disorder-averaged transmitted signal ($f(t)$) and transmission spectrum ($T$), when gradually increasing the disorder level from zero to the regime of TAI. The corresponding results are depicted in **Figure 1c**, illustrating how random disorder forces the proposed system to apply a definite first-order filtering operation to the input signal, leading to the desired output signal (marked with blue color).

To validate these findings in a full-wave 3D geometry, we map the proposed tight-binding model into a one-dimensional phononic crystal, built from a chain of acoustic quasi-bound states in the continuum (BICs)[26-29] embedded in a monomode acoustic waveguide. Such coupled bound states, resonating at the frequency of $f_0 = 2310\ Hz$, mimic the evanescently-coupled tight-binding chain described by the Hamiltonian of **Equation 1**. This leads to the realization of an acoustic topological Anderson insulator that, in the regime of dominating randomness, supports zero-energy edge modes. Note that, to the best of our knowledge, the realization of topological Anderson insulator phase in acoustics has not been reported elsewhere. In order to probe the proposed system with far-field scattering tests based on the waveguide mode, we make the radiative quality factor of the bound states finite by slightly breaking the inversion symmetry of the structure with respect to its longitudinal axis.

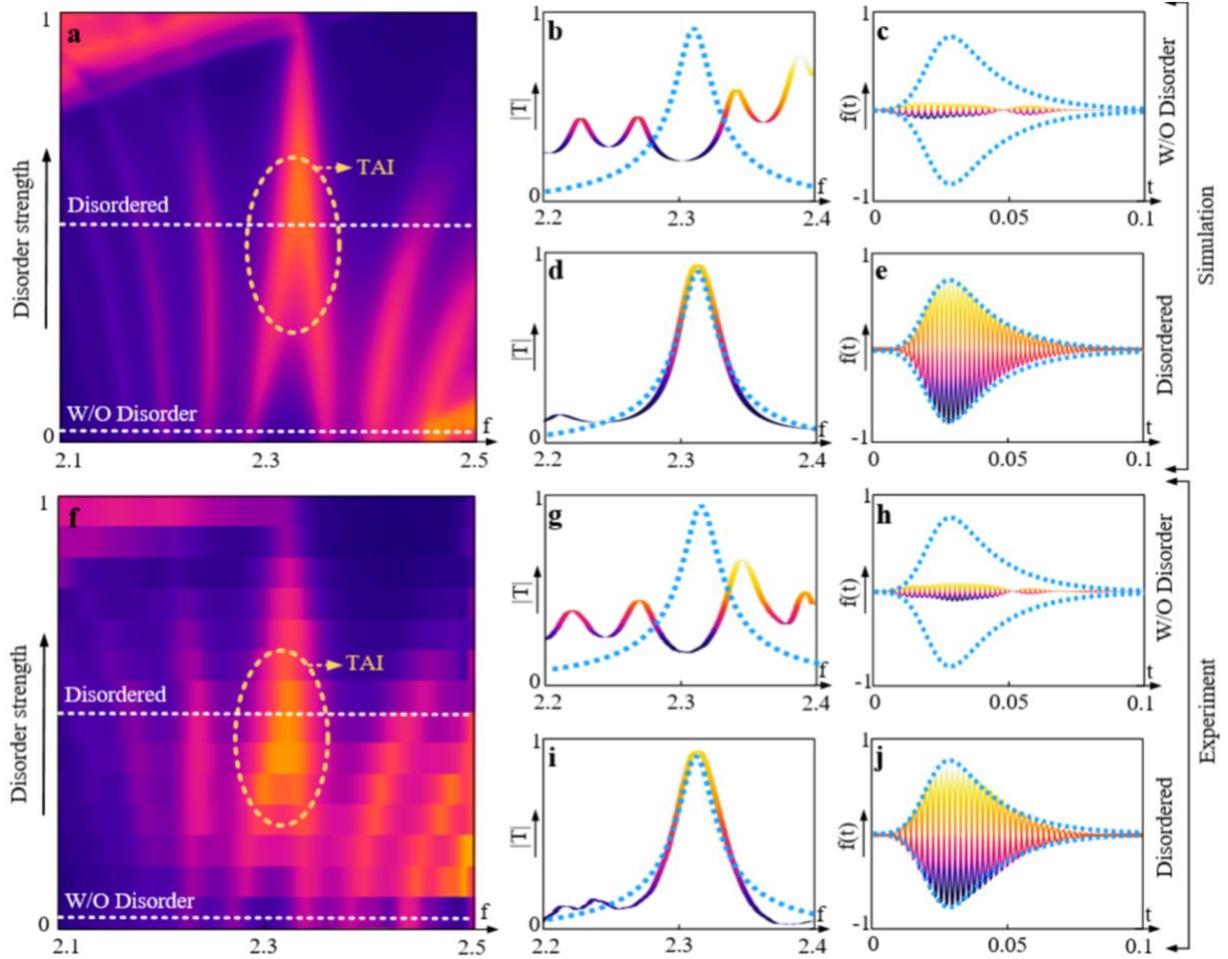

**Figure 2. Experimental demonstration of acoustic topological Anderson insulators and disorder-induced analog filtering,** We map the proposed tight-binding toy model into a 1D sonic crystal, based on coupled acoustic quasi-bound states in the continuum. **a,** Evolution of the (averaged) transmission coefficient of the phononic crystal as the strength of disorder is increased, obtained via 3D full-wave numerical simulations. The emergence of a disorder-induced zero-energy state is clear in the disorder-averaged transmission spectrum (oval region), allowing one to perform a well-defined first-order filtering operation. **b,c,** Disorder-averaged transmission spectrum and the corresponding transmitted field (numerical simulations), when the disordered-free system is excited with a Gaussian time-modulated signal. **d,e,** Same as b and c except that the system is sufficiently disordered, so that it finds itself in the TAI regime. **f,g,h,i,j,** Experimental measurements corresponding to the simulation results.

**Figure 2a** represents the transmission coefficient (averaged over disorder realizations) as a function of both frequency and disorder strength, obtained via 3D full-wave numerical simulations based on the finite element method. The result of this figure confirms the emergence of a disorder-induced resonance peak, corresponding to the zero-energy state of the TAI phase (the oval region). The possibility to leverage the Lorentzian line shape of this resonance for carrying out disorder-induced filtering is demonstrated in **Figures 2b-e**, where we have reported the averaged transmittance of the system both in the clean limit and in the topological Anderson phase. When no disorder is imparted to the system, the transmission spectrum (**Figure 2b**) exhibits a minimum due to a band gap around $f_0$, leading to an output signal (**Figure 2c**) that has an almost zero amplitude (the blue curve, which is analytically predicted). In the regime of TAI phase, on the contrary, the transmission spectrum matches the desired transfer function $H(f)$ (**Figure 2d**), corresponding to a first-order analog filter. As such, the corresponding transmitted signal, shown **Figure 2e**, follows well the targeted filtered signal (the blue curve). The experimental measurements corresponding to these numerical findings, provided respectively in **Figures 2f-j**, are in perfect agreement with the simulations. Importantly, the reported results are averaged over multiple realizations of disorder, and the standard deviation, which is inversely proportional to the system size, is kept at a small level. Note that, in principle, it is possible to achieve more complex higher-order filtering operations by constructing a network of several realizations of the proposed first-order filter (see Note IV, Supporting information).

The proposed disorder-induced acoustic filter can be used to manipulate the spectral characteristics of two-dimensional signals, i.e. images, as well. Consider the image of the Eiffel tower, shown in **Figure 3a** (top). Suppose that the pixels of this image are processed by the inverse of the target filtering function $H(f)$, therefore encrypting the image. We excite the proposed system with a signal corresponding to the encrypted image. Since, in the regime of TAI, the transfer function of the proposed system is approximately equal to $H(f)$, the system is expected to decrypt the encoded image. This is demonstrated in the insets of **Figure 3b**, illustrating how the encrypted image signal is gradually decoded by the proposed topological Anderson system, when more and more disorder is introduced. The associated experimental results, shown in **Figure 3c**, are in full agreement with the simulations. Note that decoding the image is not a trivial task as one needs not only to tune the disorder strength to the right level, but also to have some information about the required disorder statistics, in particular the difference in the standard deviations of the couplings. We note that the proposed system can be used for image edge detection as well (see Note IV, Supporting information).

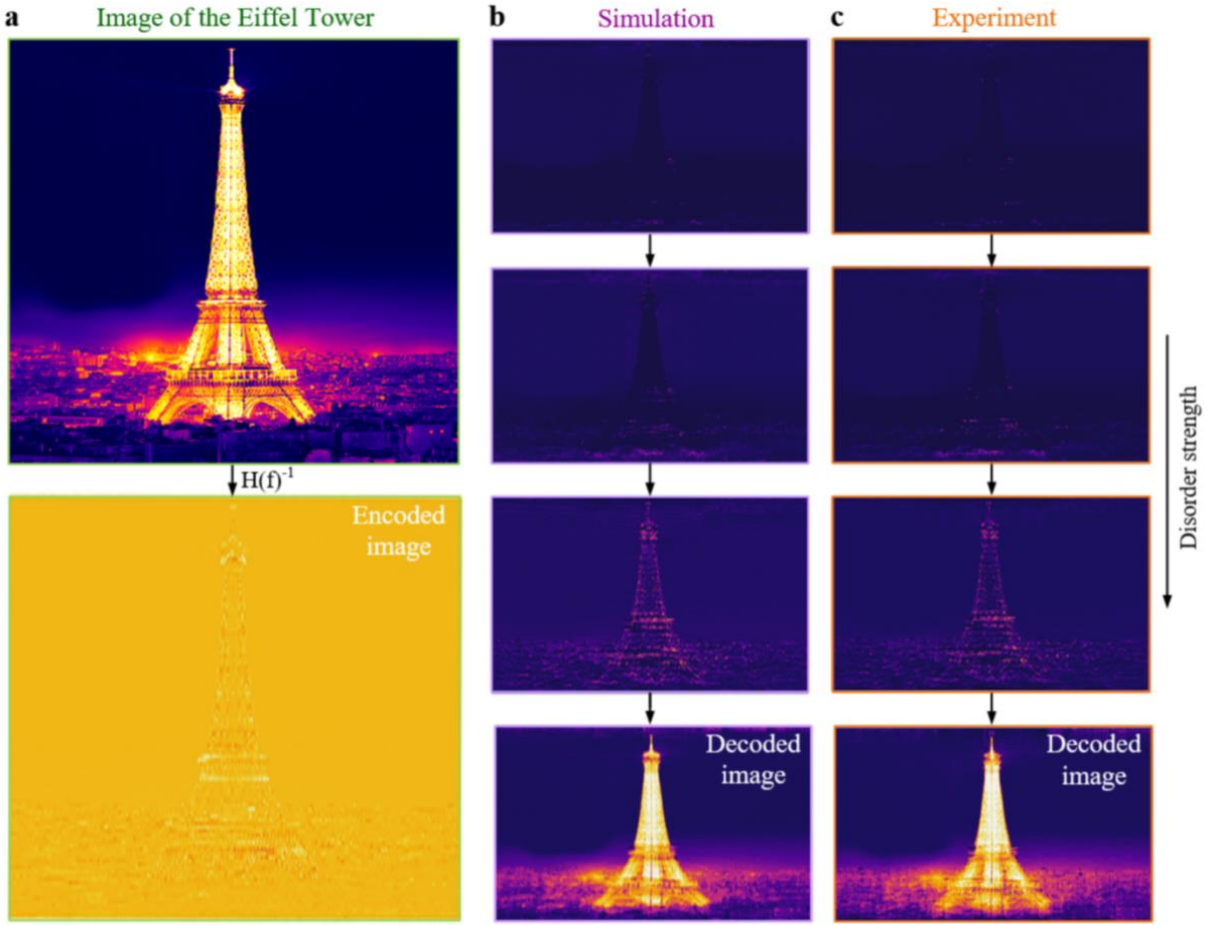

**Figure 3: Disorder-induced analog filtering of an image, a,** (Top) Image of the Eiffel tower, (Bottom) The original image is encrypted with the inverse of the target filtering function $H(f)$. The encrypted image is then fed into the input of the proposed topological Anderson system. **b,** Corresponding output images as the level of disorder is gradually increased. Since in the regime of TAI, the transfer function of the system is approximately equal to $H(f)$, the corresponding output is nothing but the decrypted image. **c,** Corresponding experimental results. The results, in agreement with numerical simulations, demonstrate the intriguing possibility of decoding the encrypted image by providing the system with more and more disorder.

In summary, we demonstrated a novel scheme for analog signal filtering, in which random disorder drives the system, forcing it to realize the desired filtering operation. Although demonstrated here in the context of acoustics, we expect the emergence of such spectral filters to be generic. In Note V, Supporting information, we demonstrate a photonic instance. Not only do our findings serve as the first experimental observation of topological Anderson insulator phase in acoustics, but also they open a new horizon for realizing a new generation of analog wave filters that rely on disorder. This is in sharp contrast to ordinary trivial analog filters in which disorder is highly detrimental (see Note VII, Supporting information).

**Experimental Section**

**Figure S1** shows a photograph of the fabricated prototype of the proposed topological sonic crystal. The sample includes a transparent pipe with square transverse cross-section, serving as a waveguide, and a chain of scatterers, that are embedded inside the waveguide. The scatterers used in the sample are commercially available Nylon 6 cast plastic rods. The width, height, and length of the waveguide are $W = 7\ cm, H = 7\ cm, L = 2\ m$, respectively, the radii of the cylinders are $R = 1.75\ cm$, the lattice constant is $a = 16.6\ cm$ and the detuning parameter $d$ is $d = 7.8\ cm$. The structure is tested in the experimental setup shown in **Figure S2.** Apart from the fabricated prototype, the setup includes three PCB 130F20 ICP® microphones, measuring the associated pressure field, a loudspeaker, generating sound and exciting the system, an acoustic Quattro Data physic analyzer, recording the associated measured data, and a computer, controlling the setup. Note also that, in order to avoid unwanted reflection and refraction, the end of the system is terminated with an anechoic termination, made of adiabatically tapered foam, shown in the bottom panel of the figure. In order to extract the disorder-averaged transmission coefficient, we excited the system with the loudspeaker, and extracted the corresponding transmission spectrum for each realization of disorder by standard standing wave pattern analysis. Then, we took the average of 10 different independent measurements, each of which corresponds to a distinct disorder configuration.

**Acknowledgements**

This work was supported by the Swiss National Science Foundation (SNSF) under Grant No. 172487.

# Supporting Information

## Note I: Topological invariant of the proposed disordered system

In this section, we investigate the topological invariant of the proposed one-dimensional system, and the locus of the topological phase transition based on the divergence of the localization length[1]. Consider again the tight-binding Hamiltonian of the proposed system, given in **Equation 1** of the main text. We define the ratio between the intra-cell and extra coefficients as the parameter $m = K_0/J_0$. The localization length of the proposed disordered system can be calculated as[1]

$$L_{local} = Abs\left( ln\left( \frac{|2(m-D_s)|^{\frac{m}{2D_s}-0.5}|2+D_s|^{\frac{1}{D_s}+0.5}}{|2(m+D_s)|^{\frac{m}{2W}+0.5}|2-D_s|^{\frac{1}{D_s}-0.5}} \right) \right)^{-1} \quad (S1)$$

in which Abs stands for absolute value, and $D_s$ is disorder strength. The topological phase transition occurs where the localization length diverges. **Figure S3a** represents the evolution of $L_{loc}$ as a function of disorder strength ($D_S$) and the parameter $m$, highlighting two distinct disconnected regions, which are characterized by different topological invariants. For our case study, $m = 0.1/0.09$=1.11. **Figure S3b** shows the corresponding winding number ($W$) for this specific $m$ as a function of disorder strength ($D_S$). As observed, $D_1 < D_s < D_2$ (TAI regime in **Figure 1**) corresponds to a non-zero topological invariant. On the other hand, $D_S < D_1$ (regime of weak disorder) and $D_s > D_2$ (regime of Anderson localization) correspond to a zero topological invariant. The values of $D_1$ and $D_2$ are $D_1 = 1.68$, and $D_2 = 3.64$.

## Note II: Numerical methods

The results of **Figure 1** of the manuscript were obtained by considering a one-dimensional periodic array (see **Figure S4**), consisting of 200 resonators evanescently coupled to each other. The parameters $\omega_0, J$, and $K$ are chosen as $\omega_0 = 1, K = K_0(1 + 0.5D_sW), J = J_0(1 +$ $D_s W$), in which $K_0 = 0.1, J_0 = 0.09$, $D_s$ is a parameter quantifying the strength of disorder, and $W$ is a random number. The chain is coupled (with the decay rate of $\gamma = 0.01\,K_0$) to two single-mode waveguides on its left and right ends. The disorder-averaged transmission spectrum was obtained by calculating the transmission coefficient through the system (using standard coupled mode analysis) for different frequencies, averaged over 20 different realizations of disorder. For the case of weak disorder, the transmission spectrum of the system shows a band gap. By increasing the disorder strength, however, the band gap closes and re-opens as topological. Considering the large number of resonators involved in chain, the two corresponding edge modes are effectively decoupled from each other, yielding a unique Lorentzian spectral line-shape around $\omega_0$ (there is effectively no mode splitting between the modes at both edges).

In order to demonstrate disorder-induced filtering operation based on such a configuration (results of **Figure 1c**), we considered, as the input, a sinusoidal signal with the frequency of $\omega_0$ (frequency of the zero-energy state of the topological Anderson phase), modulated with a Gaussian function with a variance of $\sigma = 0.1\omega_0$. The corresponding output signals were obtained by taking the inverse Fourier transform of the disordered-averaged transmission spectrum multiplied by the Fourier transform of $g(t)$.

In order to implement the proposed tight-binding toy model in an acoustic platform, we mapped it into the phononic crystal shown in **Figure S5**. The crystal consists of a conventional acoustic waveguide, surrounded with acoustic hard walls, and an array of cylinders. The width, height and length of the waveguide are $W = 7\,cm, H = 7cm, L = 2\,m$, respectively. The array is composed of 14 cylinders whose radii are $R = 1.75\,cm$. The parameters $a$ and $d$ are $a = 16.6\,cm, d = 7.8\,cm$. Each cylinder of the crystal supports a bound state in the continuum (BIC), which is odd-symmetric with respect to the vertical symmetry plane of the structure. As its name suggests, such state is perfectly bounded, while it co-exists in the radiation continuum of the waveguide. The couplings between the BIC modes can be regulated by changing the successive distance between the nearest-neighboring sites.

In order to obtain the averaged transmission spectrum, we used Comsol Multiphysics, acoustic module. The waveguide was modeled with a rectangle surrounded by walls with Neumann boundary condition ($\frac{\partial P}{\partial n} = 0$). This boundary condition was also applied to the external boundaries of the cylinders. The two lateral sides of the waveguide were surrounded by radiation boundary conditions, one of which was excited with a plane wave. The transmission spectrum, for each realization of disorder, was obtained by performing regular standing wave pattern analysis. The disorder-averaged transmission spectrum was obtained by repeating this procedure for ten different realizations of disorder and taking the average of the corresponding transmission spectra. Note that, in order to be able to detect the BIC modes in such a far-field scattering test, one has to slightly shift (6 *mm* shift in our case study) the obstacles away from the centerline of the waveguide, breaking the mirror symmetry of the structure, and allowing the BICs to leak a little bit into the continuum (this is called a quasi BICs).

**Note III: Studying the response of the proposed disorder-induced acoustic filter to other kinds of inputs**

Here, we investigate the response of the proposed acoustic system, when it is excited with input signals other than a typical Gaussian pulse. To this end, we consider an irregularly shaped pulse, modulated with the frequency of $f_0$ (**Figure S6b**). **Figure S6c** shows the corresponding (averaged) output signal, when no disorder is introduced to the system. It is seen that the output signal has nothing to do with the targeted one. Next, we suppose that the system is strongly disordered so that it enters into the topological Anderson regime. **Figure S6d** represents the corresponding output signal. It is seen that, this time, the output signal follows perfectly the solution of the targeted signal. The corresponding experimental results are shown in **Figure S7a-d**.

We also examine the functionality of the proposed disorder-induced acoustic filter for incident images other than the Eiffel tower image. Consider, for example, the photograph shown in **Figure S8a**, which is an image taken in our campus. We encrypt this image with the inverse of the target transfer function $H(f)$. The corresponding encoded image is represented in **Figure S8b**. Next, we study the evolution of the corresponding output image, when steadily increasing the strength of disorder from zero to the regime of topological Anderson localization. The corresponding (experimental) results are plotted in **Figure S8c**, constituting an evidence of the fact that the proposed system is capable of decoding the encrypted image for proper type and amount of disorder.

**Note IV: Performing other kinds of filtering operations with the proposed topological Anderson insulator system**

In this section, we discuss the possibility of performing other types of filtering operations using the proposed disorder-induced acoustic system. We start with describing how a *nth* order analog filter can be realized by constructing a network of the proposed disordered system.

In frequency domain, the transfer function of a nth order analog filter is expressed as

$$H(\omega) = \frac{B}{(j(\omega - \omega_0)^n) + A_{n-1}(j(\omega - \omega_0))^{n-1} + .. + A_1(j(\omega - \omega_0)) + A_0} \quad (S2)$$

$H(\omega)$ can be re-written as

$$H(\omega) = \sum_{i=1}^{n} H_i(\omega); \quad H_i(\omega) = \frac{K_i}{j(\omega - \omega_0) + \frac{\omega_0}{2Q_i}} \quad (S3)$$

where $Q_i = -\omega_0/2P_i$, in which $P_i$ are complex poles of the associated nth order polynomial, and $K_i$ are the corresponding constant coefficient. Equation S3 suggests a straightforward method for realizing the nth order filtering operation. One has to add the output signals of several Lorentzian resonators with different quality factors $Q_i$. This can be accomplished in a fully analog manner as in **Figure S9**, in which the output signals of several resonators (associated with the edge modes of several acoustic topological Anderson insulators) with different quality factors are added to each other using an array of rat-race couplers[2]. Note that the quality factor of most resonators (including the one of the proposed topological Anderson metamaterial) can be readily tuned by tailoring the dissipation losses in the system.

In **Figure S10a**, we have demonstrated a second-order analog filter based on the scheme described above. In this configuration, we have added the output signals of two realizations of first-order filters based on the acoustic topological Anderson insulator phase demonstrated in the main text. Each realization corresponds to a different dissipation loss. As a result, the corresponding transfer functions $\boldsymbol{H_1(\omega) = 6/(j(\omega - \omega_0) + 18)}$ and $\boldsymbol{H_2(\omega) = 12/(j(\omega - \omega_0) + 12)}$ are different. The rat-race coupler adds these two transfer functions to each other, leading to a new transfer function $\boldsymbol{H_3(\omega) = H_1(\omega) + H_2(\omega)}$ that corresponds to a second order filter. In order to verify the functionality of the system, we consider an arbitrarily shaped signal as the excitation (**Figure S10b**), and calculate the corresponding output signal (**Figure S10c**). It is observed that the output of the system matches the targeted filtered signal.

We note that an alternative route to realize transfer functions other than Lorentzian is to cascade two or several topological Anderson insulators, allowing their edge modes to couple to each other. The constructive and destructive interference between the associated edge modes may create a Fano-like spectra (rather than Lorentzian) in the regime of strong randomness, opening the way for more complex and sharper transfer functions including comb-like spectra.

While in the main text we focused our attention on the transmission of the proposed system, the corresponding reflected signal is relevant for another kind of filtering operation, namely band-stop filtering. As explained in the main text, near the resonance frequency of the edge mode of the topological Anderson insulator phase, there exists a resonance peak in the averaged transmission spectrum. This corresponds to a dip (zero) of the reflection spectrum (**Figure S11a**). Near this zero, assuming that the input signal has a sufficiently small bandwidth, the reflection coefficient of the structure can be approximated with a linear function of the form $R(\omega) \approx A\omega$ (see the dashed blue line in Figure S9a). Interestingly, $R(\omega)$ is similar to the transfer function of a differentiator filter. As a result, for the input signals that are sufficiently wide in temporal domain, the reflected signal is nothing but the derivative of the incident field. In order to demonstrate this, we consider again our tight-binding toy model and assume that it is excited with a Gaussian-type input signal (**Figure S11b**). The corresponding reflected signal is indicated in **Figure S11c**. It is observed that the reflected signal has a Gaussian derivative profile.

One common technique to detect the outlines of an image is to use a differentiation filter. We have shown that our proposed disorder-induced system can act as such a filter. As such, it is capable of resolving the edges of incident image (in one direction). In order to demonstrate this possibility, we assume that the input signal is modulated with the image shown in **Figure S11d** (top). The corresponding reflected image is shown in **Figure S11d** (bottom). It is observed that the vertical edges of the image are resolved, confirming the excellent performance of the proposed filter.

## Note V: Photonic topological random filters

Here we explore the possibility of generalizing the proposed disorder-induced topological acoustic filter to electromagnetics. To this end, similar to the acoustic case, we consider a regular photonic crystal consisting of a conventional metallic waveguide. In Comsol Multiphysics, we model the waveguide via a rectangular pipe surrounded with perfect electric conductors. On the other hand, we model the rods with cylinders filled with a material (silicon) with the refractive index of 3.4. We then introduce some disorder into the successive distance between the inclusions of the photonic crystal. In the regime where the strength of disorder is weak, the transmission spectrum is gapped around the resonance frequency of the resonators, due to the topologically trivial nature of the system in the clean limit. Adding disorder to the system, however, closes the trivial gap and re-opens it as topological, as broadly discussed in the main text. This gives rise to a zero-energy edge state, manifesting itself as a Lorentzian resonance peak in the averaged transmission spectrum, which is employed to perform first-order band-pass filtering. To demonstrate these predictions, we consider a Gaussian-modulated sinusoidal signal (with frequency of $\boldsymbol{f_0}$ and the variance of $\boldsymbol{\sigma = 0.1 f_0}$) as the input signal. We then study what the system returns both in the clean and TAI regimes. **Figures S12a,b** show the output signals corresponding to the clean and disordered systems, respectively. When the system is free of disorder (panels a), the output signal has a very low level, being far from the targeted filtered signal (blue dashed line). On the other hand, the output signal of the disordered system (panel b) is in perfect agreement with the targeted signal.

**Note VI: Analysis of the proposed disorder-induced system based on Finite difference time domain (FDTD) approach**

In order to obtain the signal coming out of the proposed system, we calculated the inverse Fourier transform of the disorder-averaged transmission coefficient multiplied by the Fourier transform of the input signal. The alternative, more straightforward, strategy is to calculate the output signal via direct FDTD simulations. To do that, we suppose that the system is excited with a Gaussian-modulated sinusoidal and study the corresponding averaged output. In **Figures S13**, we plot the corresponding output signals, averaged over 10 different realizations disorder, for both disorder-free and disordered systems. The results of these figures are in complete agreement with the ones of **Figure 2a** of the main text.

**Note VII: Studying the effect of disorder on trivial analog wave filters**

In this section, we compare our proposed disorder-induced topological analog filter with a trivial ordinary one. Consider, for example, the resonant structure shown in **Figure S14**, based on defect tunneling through a Bragg band gap. Such a structure is also capable of performing first-order filtering operation (see the top signal path). However, when some disorder is added to the crystal, the transfer function of the system and the corresponding output significantly deviate from their original cases, as it is confirmed by direct numerical simulation. The behavior of our proposed resonant structure, based on TAI phase, is drastically different. Not only is disorder not detrimental to our proposed system, but, on the contrary, it is necessary, triggering the functionality of interest.

**Note VIII: Supporting Figures**

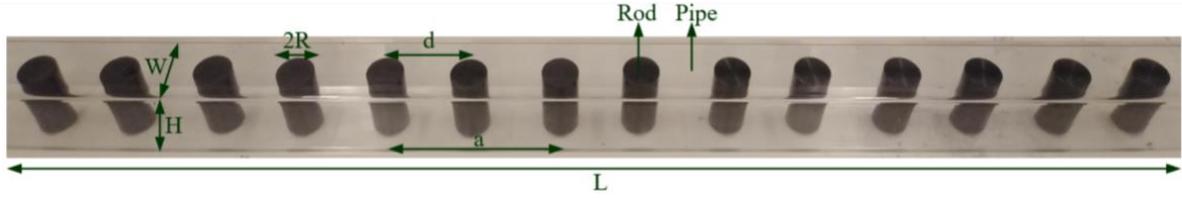

**Figure S1. Fabricated prototype of the phononic crystal,** The sample consists of a rectangular pipe, taking the role of the acoustic waveguide, and a set of nylon cast plastic rods embedded inside the waveguide.

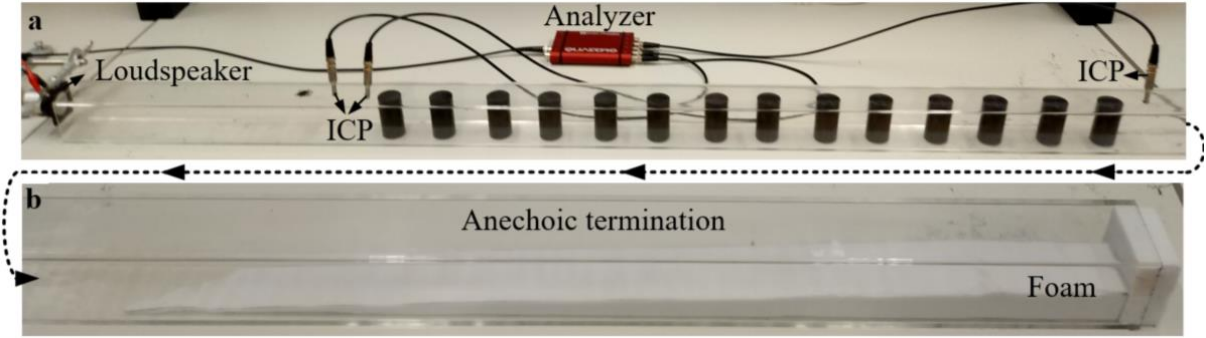

**Figure S2. Experimental setup used to test the proposed acoustic topological Anderson filter,** In addition to the fabricated sample, the setup consists of an acoustic Quattro Data Physics analyzer, three ICP® microphones, a loudspeaker and an acoustic termination, made from appropriately tapered foam.

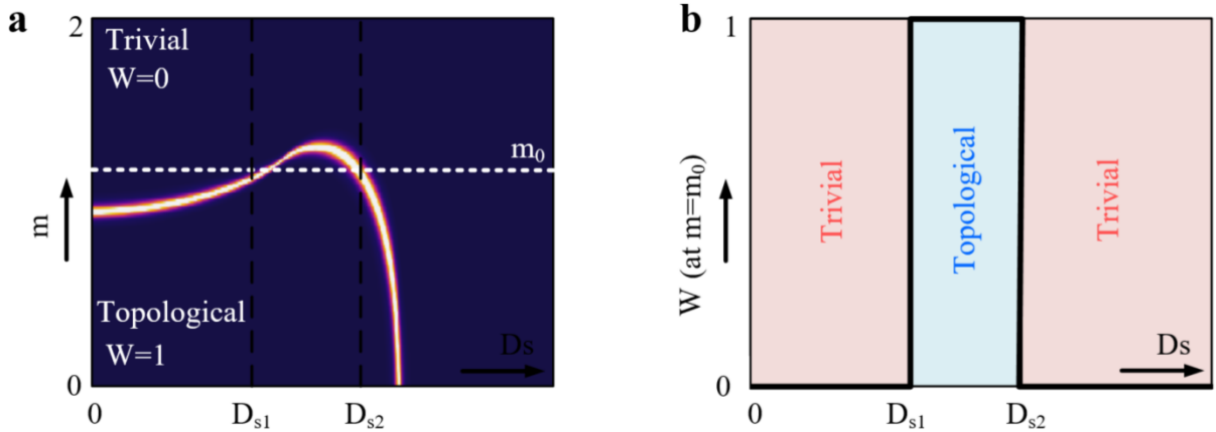

**Figure S3. Topological invariant of the proposed system**, **a,** Localization length of the proposed disordered system as a function of disorder strength ($D_s$) and the parameter $m$. **b,** Calculated winding number at the specific value of $m = 1.11$ corresponding to our experiment.

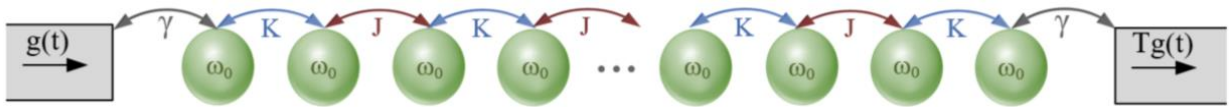

**Figure S4. Tight-binding toy model used to perform disordered-induced topological filtering**, The model consists of an array of 200 resonators, evanescently coupled to each other with specified coupling coefficients and on-site potentials. From its right and left corners, the chain is coupled to two single mode waveguides with the tunneling rate of $\gamma$.

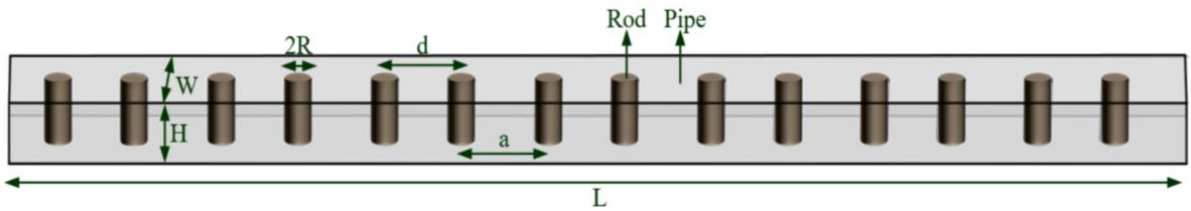

**Figure S5. Phononic crystal used to perform disorder-induced topological acoustic filtering,** The crystal is composed of an acoustic waveguide and an array of cylinders with dimensions specified in the text. Each cylinder supports a resonance state that, while coexists in the radiation continuum of the waveguide, remains perfectly bounded.

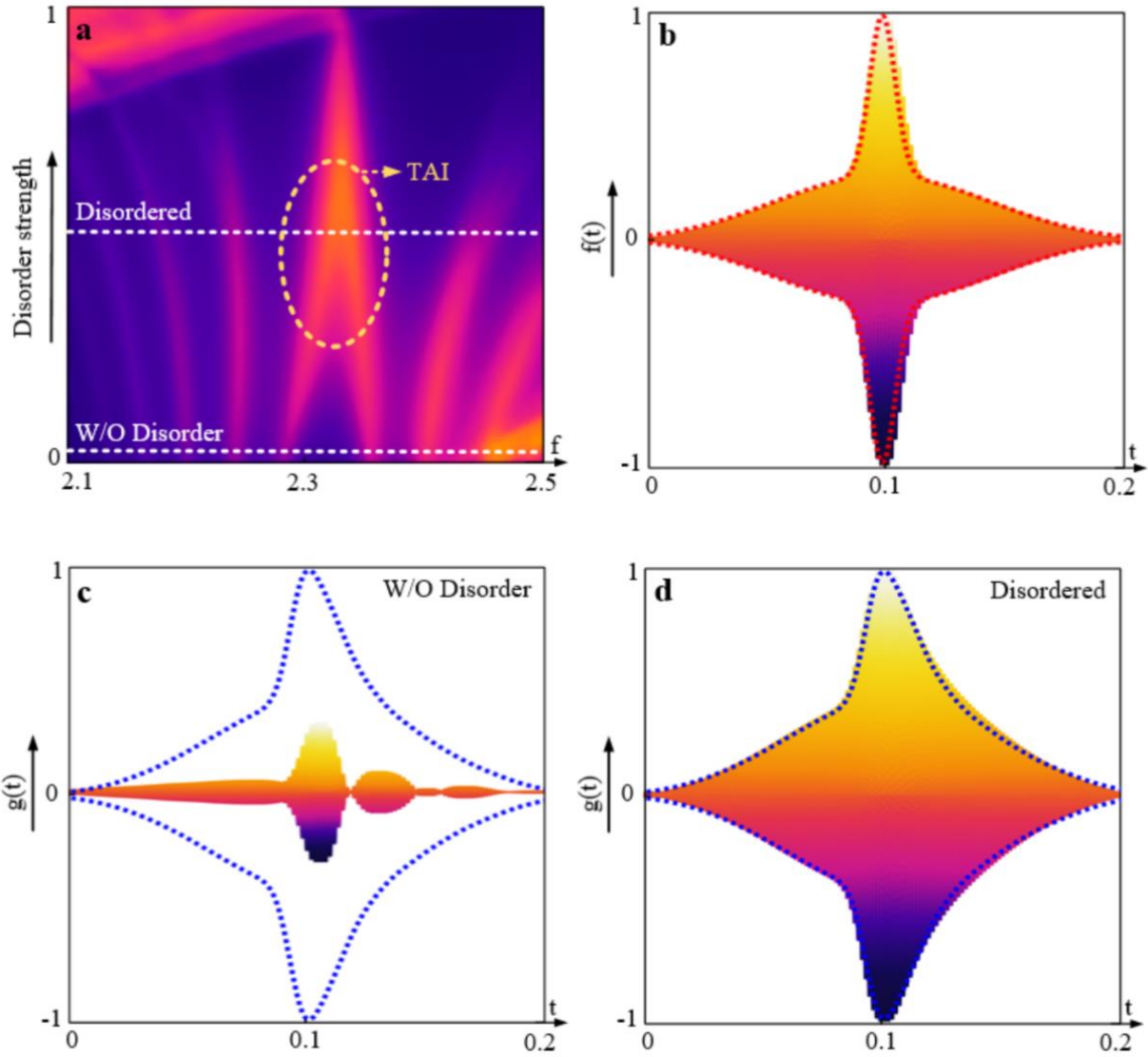

**Figure S6. Response of the proposed system to an irregularly-shaped signal, a,** Disorder-averaged transmission coefficient of the proposed system versus disorder, **b,** We suppose that the system is excited with an irregularly shaped signal shown in the inset, **c,** Corresponding output signal in the disorder-free limit, being far from the targeted filtered signal (blue line). **d,** Corresponding output in the regime of topological Anderson phase.

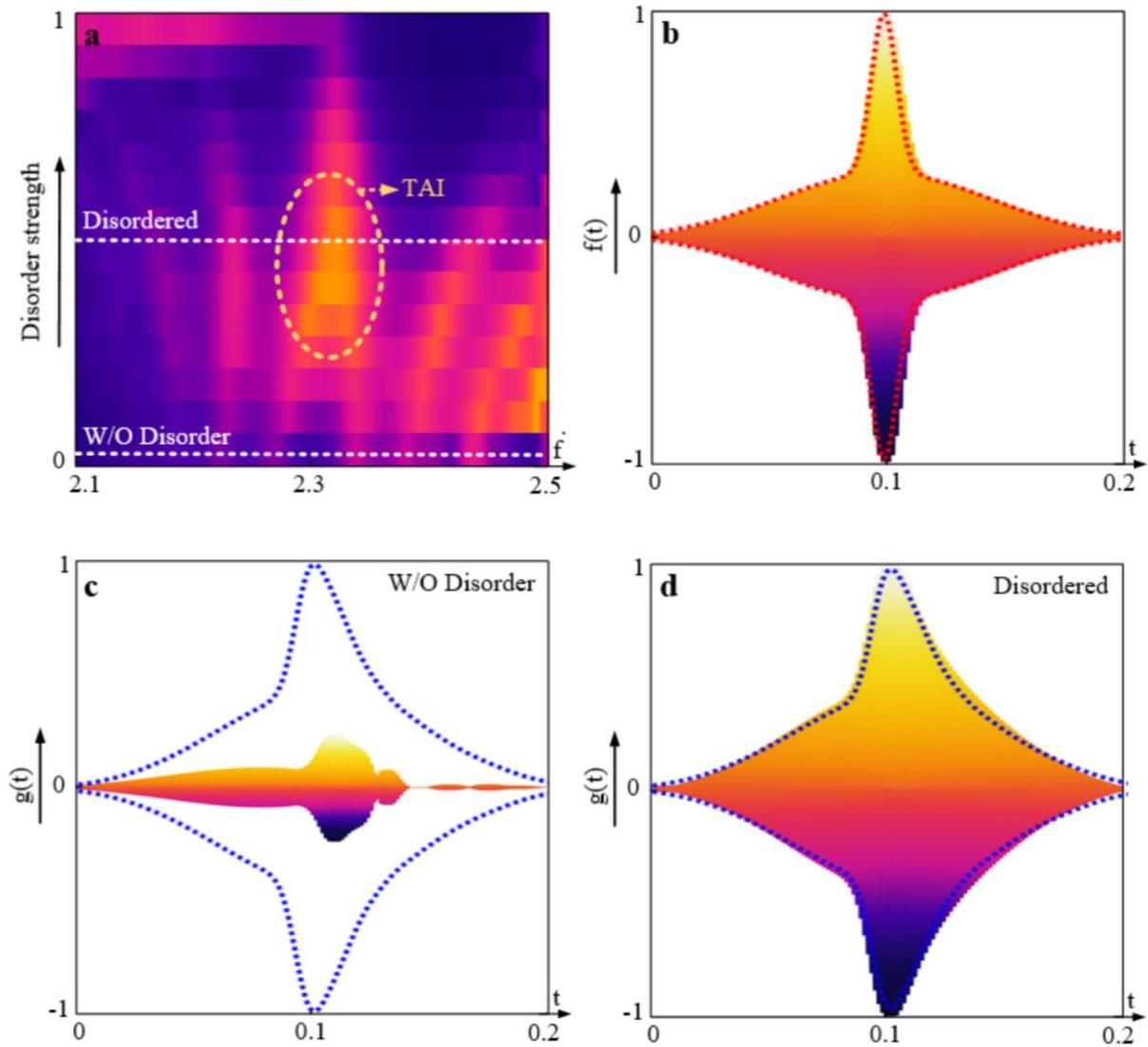

**Figure S7. Response of the proposed acoustic system to an irregularly-shaped signal,** The figure repeats the analysis of **Figure S6**, except that the results are obtained based on the measured (averaged) transmission spectrum shown in panel a.

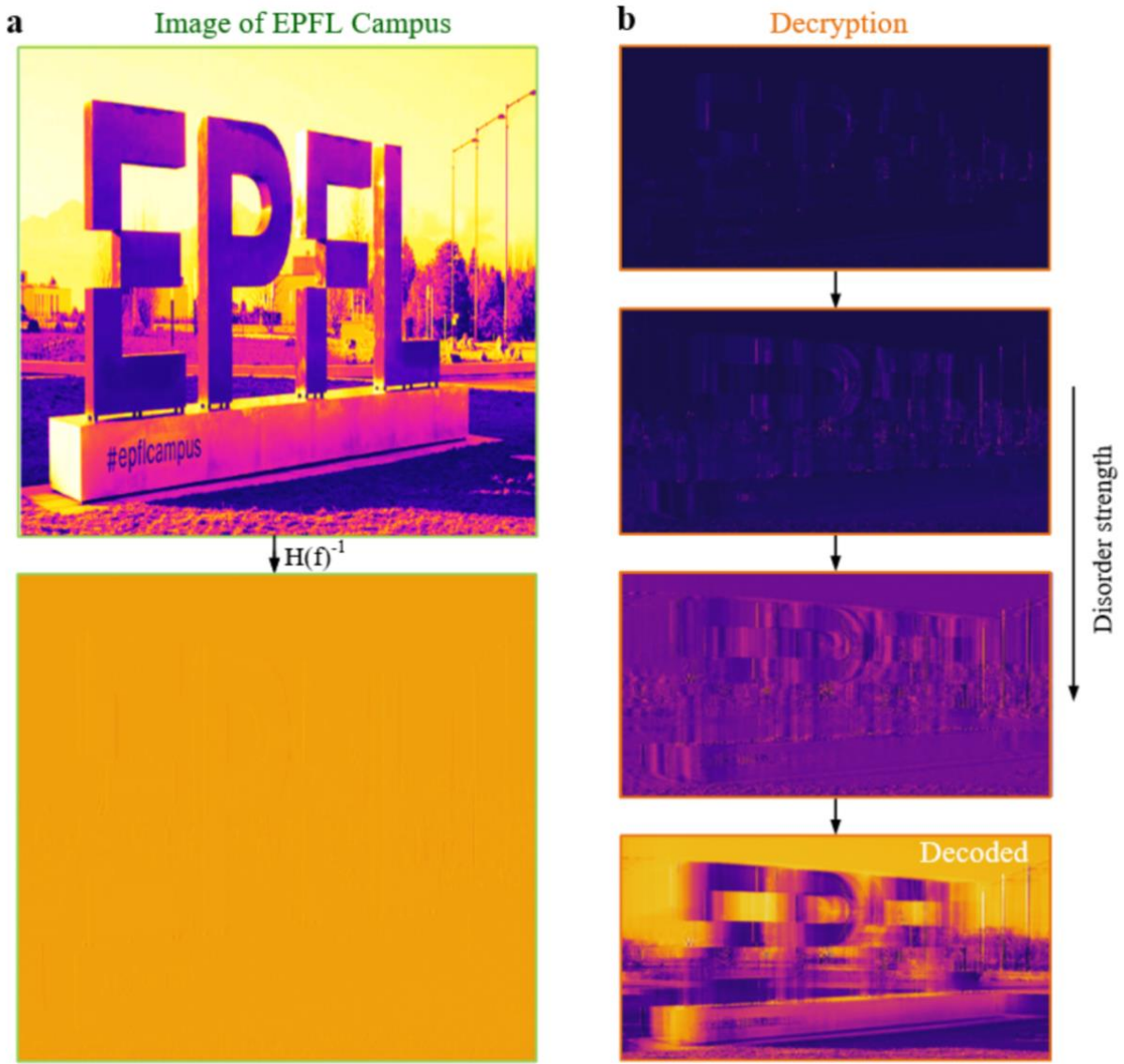

**Figure S8. Performing filtering operation in an image taken in our campus,** The figure repeats the analysis of **Figure 3** of the main text, for a different test image taken on our campus. The results are obtained based on the measured (experimental) transmission spectrum shown in **Figure S7a**.

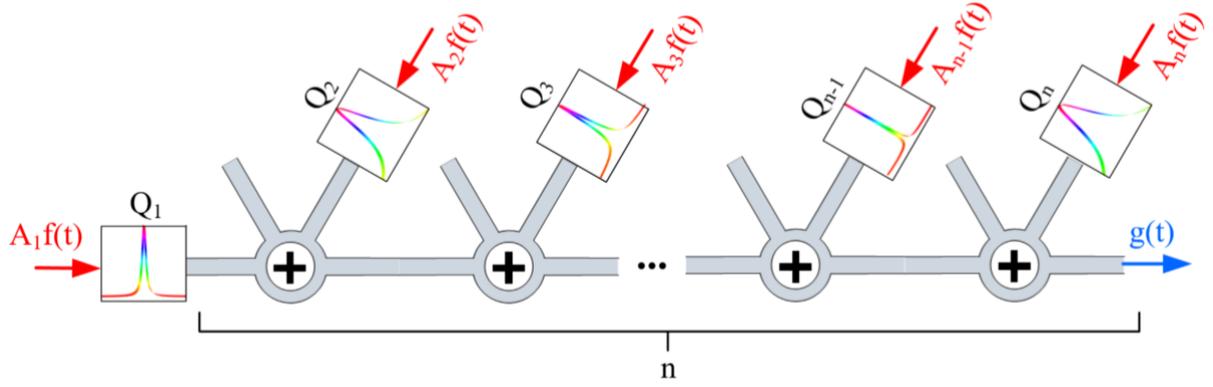

**Figure S9. Performing higher-order filtering operation by constructing a network of the proposed disorder-induced first-order filter,** The output signals of n different Lorentzian resonators (with different quality factors) are added to each other, leading to the realization of a nth order filter. The summation is accomplished by an array of acoustic rat-race couplers[2].

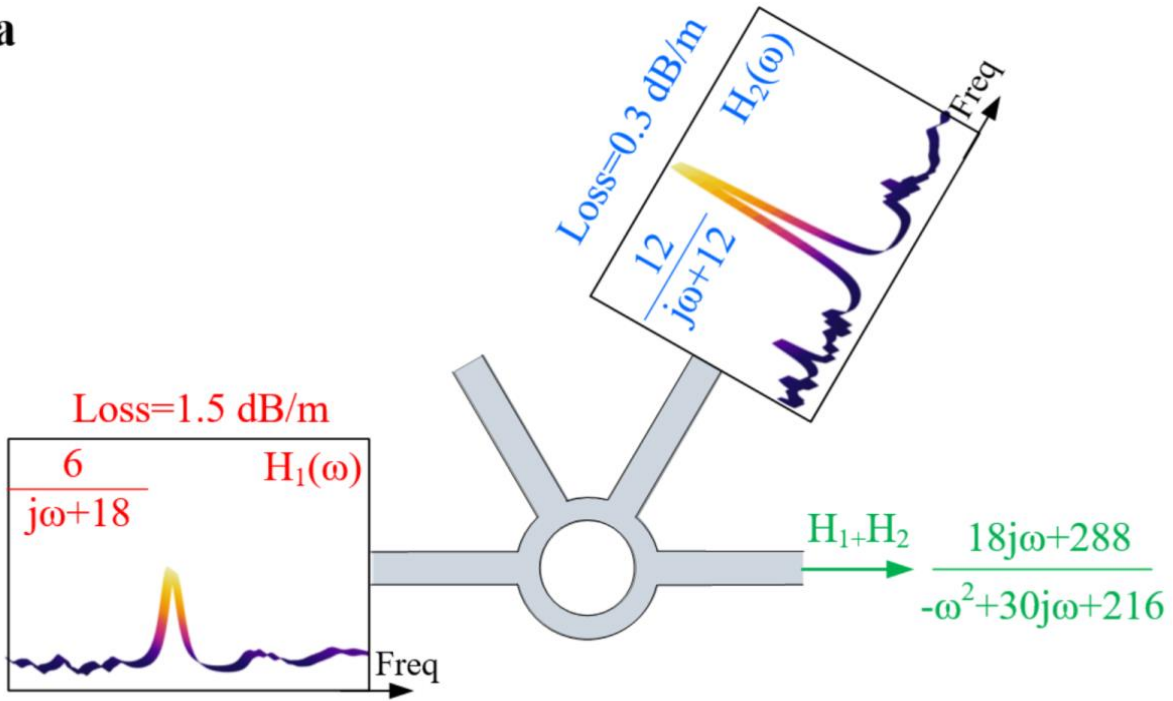
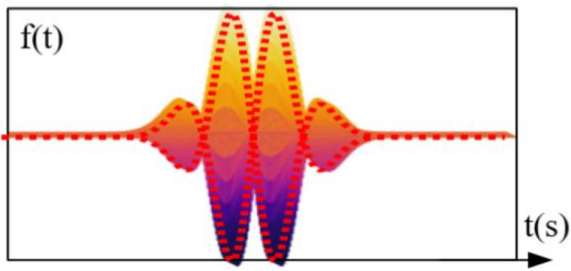
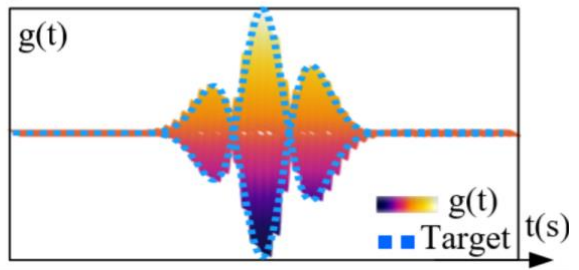

**Figure S10. Second order topological acoustic filter, a,** The output signals of two first-order disordered-induced filters are added to each other using acoustic rat-race couplers. This leads to the realization of the transfer function $H_3 = H_1 + H_2$, associated with a second-order

filter. **b,** An arbitrarily shaped signal is considered as the excitation, **c,** Corresponding output signal.

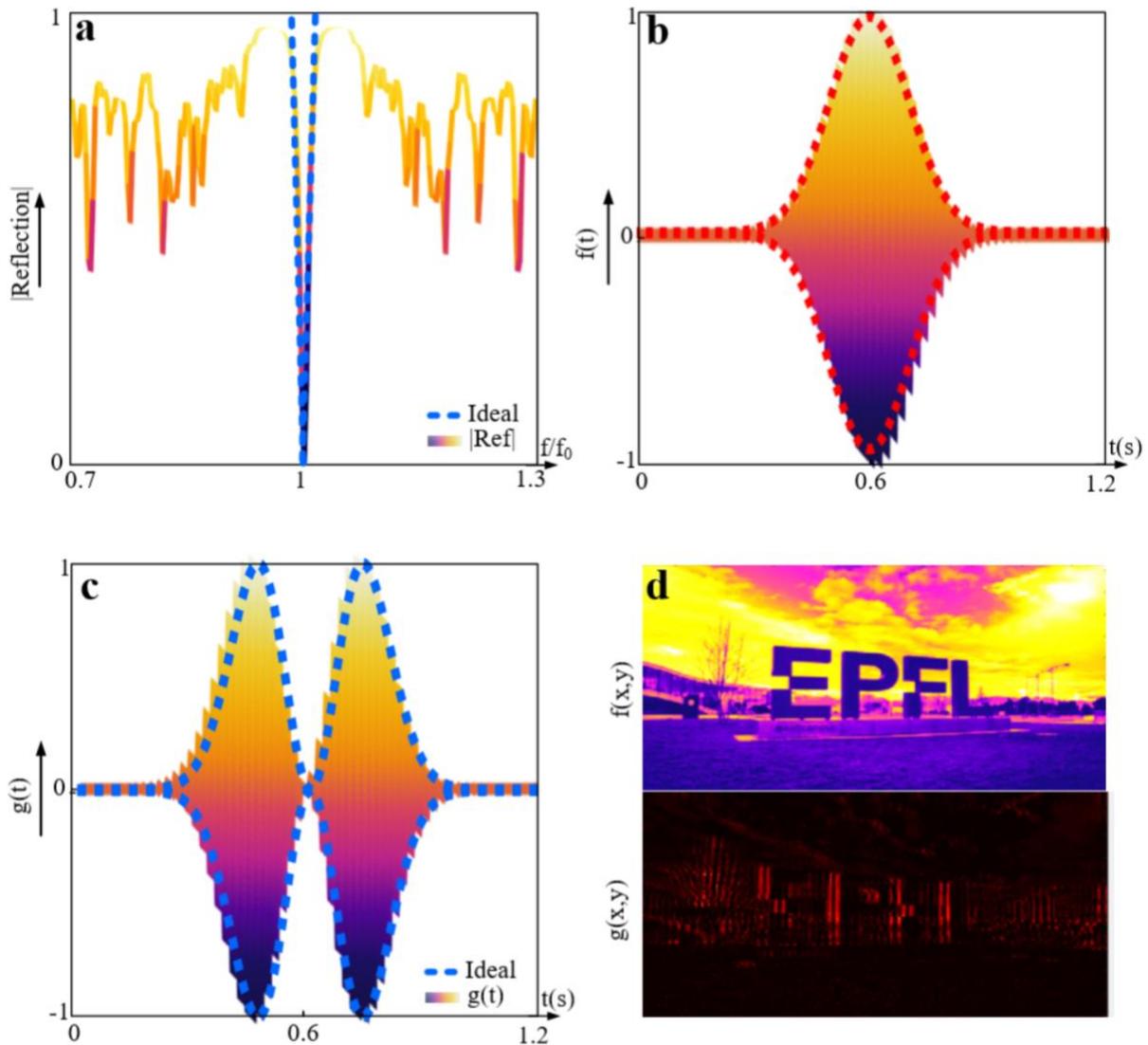

**Figure S11. Disorder-induced analog band-stop filtering, a,** Reflection coefficient of the proposed system in the regime of topological Anderson insulator phase. The spectrum exhibits a dip (zero) near the resonance frequency of the edge mode. Near this dip, the reflection coefficient can be approximated with a linear function. **b,** A Gaussian-type signal is considered as the excitation, **c,** Corresponding output, possessing a Gaussian derivative profile. **d,** Resolving the edges of an incident image modulated with an image taken our campus.

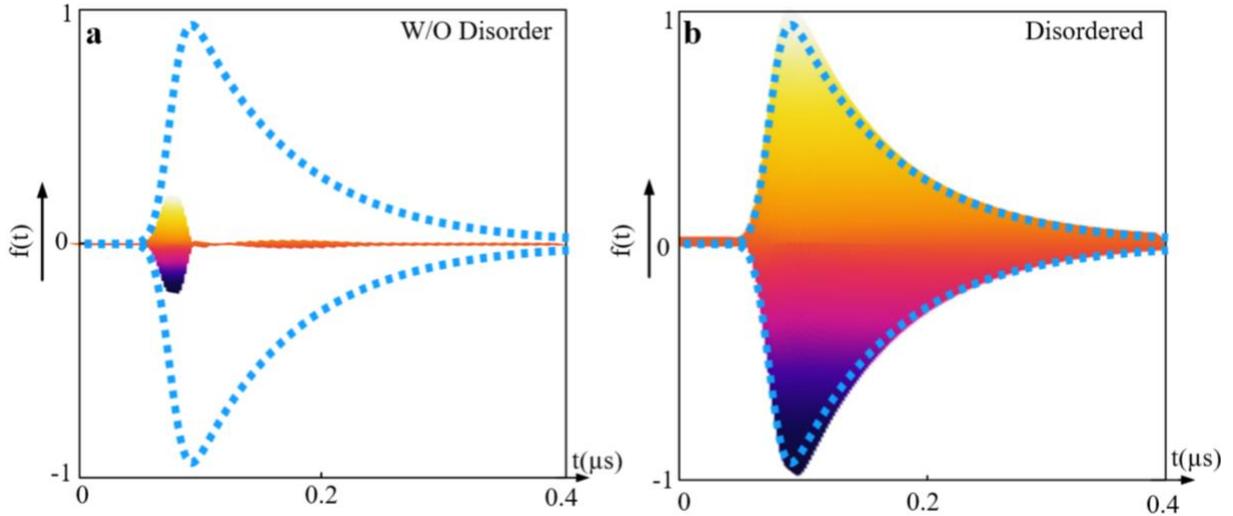

**Figure S12. Demonstration of disorder-induced filtering in electromagnetics,** We consider a photonic crystal quite similar to the phononic crystal shown in **Figure S5**, consisting of silicon rods arranged inside a conventional metallic waveguide. In the clean limit, the system is designed to be topologically trivial. Yet, introducing disorder to it enables topological phase transition, leading to topological Anderson insulator phase. **a,** Disorder-averaged transmitted signal, when the system is excited with a Gaussian-type time modulated signal and only weak amount of disorder is present. **b,** Same as a except that the system is strongly disordered so that it enters TAI regime.

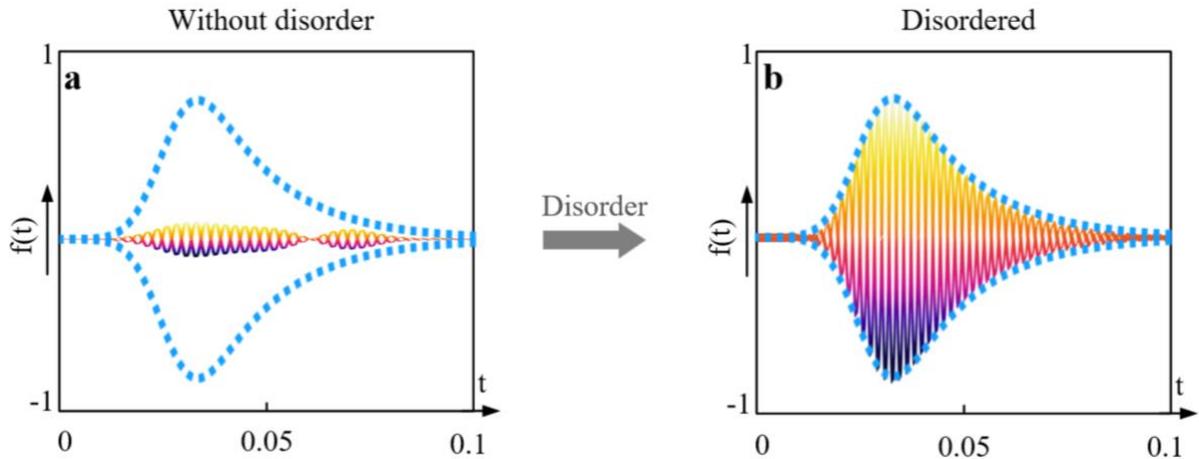

**Figure S13. FDTD analysis of the proposed acoustic system**, The figure repeats the analysis of **Figure 2** of the main text except that the results are obtained via direct finite difference time domain simulations, **a,** Corresponding output signal when no disorder is present in the system under study, **b,** Corresponding output signal when the system is strongly disordered.

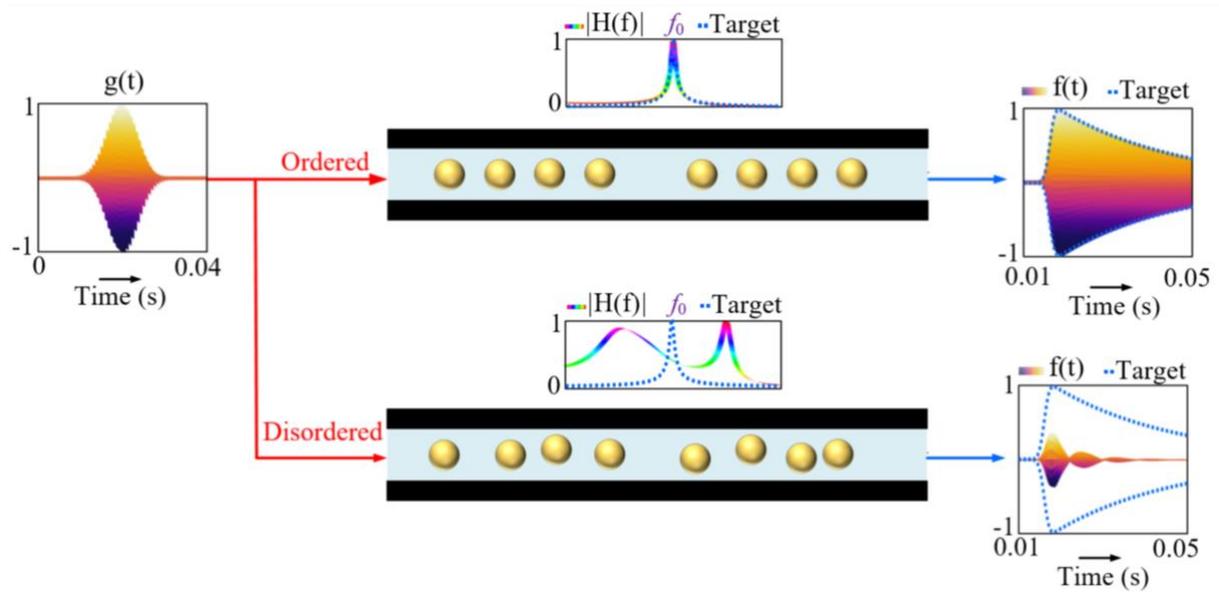

**Figure S14. Trivial analog acoustic filtering,** we consider the resonance associated with the defect tunneling through a Bragg band gap (top signal path). Near the resonance, the system acts as a first-order spectral filter. When some disorder is added to the system, the transfer function of the system and the corresponding output signal significantly deviates from the original cases.